\documentclass[aps,prl,reprint,groupedaddress,onecolumn]{revtex4-1}
\usepackage{appendix}

\usepackage{multimedia}
\usepackage{lipsum}
\newcommand{\dee}{\mbox{d}}

\newcommand{\Thio}{\textit{Thiovulum majus}}
\newcommand{\thio}{\textit{T. majus}}

\newcommand{\thin}{\theta_{\rm in}}

\usepackage{graphicx}
\usepackage{amssymb,amsfonts,amsmath,times,color}
\usepackage[english]{babel}
\begin{document}
\title{Trapping and Scattering of a Multiflagellated Bacterium by a
  Hard Surface} \author{Alexander P Petroff \& Schuyler McDonough}
\affiliation{Department of Physics, Clark University, Worcester, MA
  01610, USA} \date{\today}

\begin{abstract}
\textit{Thiovulum majus}, which is one of the fastest known bacteria,
swims using hundreds of flagella.
Unlike typical pusher cells, which swim in circular paths over hard
surfaces, a \thio~cell turns its flagella normal to the surface.
To probe the torques that stabilize this unusual fixed
point, the trajectories of several thousand collisions between a
\thio~cell and a wall of a quasi-two-dimensional microfluidic chamber
are analyzed.
Measuring the fraction of cells escaping the wall either to the left
or to the right of the point of contact---and how this probability
varies with incident angle and time spent in contact with the
surface---maps the scattering dynamics onto a first passage problem.
These measurements are compared to the prediction of a Fokker-Planck
equation to fit the angular velocity of a cell in contact with a hard
surface.
This analysis reveals a bound state with a narrow basin of attraction
in which cells orient their flagella normal to the surface.
The escape angle predicted by matching these near field dynamics with
the far field hydrodynamics is consistent with observation.
We discuss the significance of these results for the ecology of \thio~
and their self organization into active chiral crystals.

\end{abstract}

\maketitle

\section{Introduction}
Microbes in their natural environments often live in close proximity
to surfaces such as the air-water interface, sediment grains in
water-saturated soils, and sinking
detritus~\cite{fletcher2013bacterial,harshey2003bacterial,azam2001sea,datta2016microbial}.
When a cell swims near a surface, the flow it generates is perturbed
by the presence of the boundary. The perturbed flow turns and advects
the swimming cell~\cite{lauga2009hydrodynamics,lopez2014dynamics}.
Microbes have evolved to exploit these flows to better colonize new
environments~\cite{durham2012division}, improve foraging and mating
strategies~\cite{drescher2009dancing}, and navigate heterogeneous
environments~\cite{muller2020compass,waisbord2021fluidic,petroff2022phases}.

Past work on the hydrodynamic coupling between cells and surfaces has
focused on a relatively small number of domesticated
species~\cite{lauga2009hydrodynamics}, which poorly represent the true
diversity of microbial form and locomotion observed in
nature~\cite{zinder2006morphological}.
To better understand how more complex flagellation patterns allow
microbes to exploit new hydrodynamic effects, we study the sediment
microbe
\Thio~\cite{de1961observations,lariviere1963cultivation,wirsen1978physiological,marshall2012single},
which is one the the fastest-known
bacteria~\cite{schulz2001big}.
\thio~uses hundreds of flagella to swim at a speed up to
$U\approx600\,\mu$m$\,$s$^{-1}$~\cite{garcia1989rapid} and turn
smoothly in chemical gradients~\cite{thar2003bacteria}.
%
% The body is slightly prolate and quite large, having a typical
% diameter of $2a\approx7\,\mu$m and aspect ratio $1.15$. They swim
% using several hundred flagella, rather the the $1$--$10$ flagella
% typical of model organisms. Moreover these flagella are unusually
% short relative to body size.
%
\thio~cells attach to surfaces by means of a $\sim100\,\mu$m-long
mucus tether that is extruded from the cell
posterior~\cite{de1961observations}. Once attached, the cell rotates
its flagella to generate a flow that efficiently stirs its chemical
environment~\cite{fenchel1998veil,thar2002conspicuous,petroff2015biophysical}.
When many free-swimming cells collide with a flat surface, they
self-organize into two-dimensional active
crystals~\cite{petroff2015fast,petroff2018nucleation}.
The fluid mechanics that underlie the tethering of cells to surfaces
and the formation of active crystals are not well understood.
%

%  We enrich these bacteria, using previously described methods, and load
%  them into a microfluidic chamber. The thickness $125\,\mu$m of the
%  chamber is much greater than the typical size $a=3.5\pm0.5\,\mu$m of
%  the cell. We focus on a plane midway between the top and bottom of the
%  chamber and track the cells as they collide with the side wall of the
%  chamber at a random angle.

Here we experimentally investigate these dynamics by tracking the
motion of several thousand \thio~cells as they collide with the wall
of a quasi-two-dimensional microfluidic chamber.
After a collision, a cell remains in contact with the surface for an
average trapping time $\langle t \rangle =0.21\pm0.01\,$s. However
trapping times are widely distributed.
In approximately $20\%$ of collisions, the cell is only momentarily
($t<0.05\,$s) in contact with the surface before escaping back into
the bulk fluid. Fig.~\ref{entropy}(a) and supplemental video SV1.mp4
show a one such collision.
A similar fraction of collisions result in the trapping of the cell
(Fig.~\ref{entropy}[b] and supplemental video SV2.mp4) near the
surface for more than $0.5\,$s and as long as $17\,$s.
These trapped bacteria do not swim along the circular paths that are
typical of pusher cells~\cite{lauga2009hydrodynamics}. Rather, a cell
orients its flagella normal to the
surface~\cite{das2019transition,ishimoto2019bacterial,petroff2015fast}
and lateral motion is purely diffusive~\cite{petroff2018nucleation}.
The distribution of escape times for all collision is shown in
Fig.~\ref{entropy}(c).

%  The trapping of 
%  %
%  Here we seek to understand how a multiflaggellated cell can become
%  trapped by a hard surface.
%  %

Understanding the dynamics that lead a \thio~cell to either scatter
quickly from a surface or become trapped for extended periods is
challenging from both a theoretical and experimental stand point.
It is difficult to gauge the relative importance of contact forces and
hydrodynamic forces even for well characterized cells with few
flagella~\cite{lauga2009hydrodynamics,drescher2011fluid,lushi2017scattering}.
\thio~is far more complex than these model organisms.
It is covered with several hundred flagella and it remains unknown how
these flagella interact with one another (e.g., by forming bundles).
It similarly difficult to probe these dynamics experimentally.
Because \thio~cannot be grown in pure culture, the standard
experimental techniques (e.g., genetic manipulation of strains) that
have proven vital to the study of model organisms are unavailable.
Extending our understanding of cell-surface interactions from model
organisms to the broader field of environmental microbiology requires
experimental techniques and analysis to overcome these challenges.

%  material properties of these flagella such as the
%  
%  
%  Despite these challenges, the 
%  
%  
%  
%  Past work has shown that cells, such as those in
%  Fig.~\ref{entropy}(b), may become hydrodynamically trapped by the
%  surface and exert a force normal to the
%  surface~\cite{petroff2015fast}. As rotational diffusion misaligns the
%  flagella, the cell drifts slightly over the surface and eventually
%  escape~\cite{petroff2018nucleation}.
%  %
%  Theoretical analysis shows that a spherical cell with a single short
%  flagellum turns to exert a force normal to the
%  surface~\cite{das2019transition,ishimoto2019bacterial}.
% 

%  Directly modeling the forces and torques that trap and scatter a
%  swimming cell is impractical as each cell is covered in hundreds of
%  flagella which are moved by both
%  hydrodynamic~\cite{} and contact
%  forces.
%  %

We probe the dynamics of scattering and trapping indirectly by
analyzing the statistics of 3661 collisions between a swimming cell
and a wall.
We observe striking simplicity in the trajectories of escaping cells.
Although the distribution of incident angles is widely distributed
(Fig.~\ref{entropy}[d]), escape angles are narrowly distributed
(Fig.~\ref{entropy}[e]).
Cells that are in fleeting ($t<0.1\,$s) contact with the surface tend
to escape at an average angle of $69.2^\circ\pm0.4^\circ$
Any such cell that approaches the wall from the left escapes to the
right and \textit{vice versa}. In our coordinate system, these cells
escape at an angle $\theta_{\rm out}>0$.
By contrast, cells that remain in contact with the surface for at least
$0.5\,$s are bimodally distributed, being equally likely to escape at
either positive or negative angles with mean $\langle|\theta_{\rm
  out}|\rangle=58.0^\circ \pm 0.5^\circ$.

The bimodality of escape directions develops gradually as a cell
remains in contact with surface.
We measure the probability $P_{+}$ that a cell escapes the surface at
an angle $\theta_{\rm out}>0$ and calculate the the associated
binomial entropy $S$ of the escape direction.
As shown in Fig.~\ref{entropy}(f), the information of a cell's
trajectory before the collision is gradually lost over the course of
half a second.

In this article, we show how this erasure of information can be mapped
onto a first passage problem~\cite{berg1993random,moen2022trapping} to
infer the torques acting on a multiflagellated bacterium that is in
contact with a wall. Matching the inferred near field dynamics with the
far field hydrodynamics of a pusher cell predicts the angle at which
cells escape.

%  Remarkably, all cells escape the surface at similar angles, having a
%  magnitude between  (see
%  Fig.~\ref{entropy}[c]). However, the direction of escape differs
%  between the two types of collisions.  By contrast, cells are equally likely to
%  escape in either direction.
%  
%  
%  
%  This memory is gradually erased as the cell remains in contact with
%  the surface.
%  %
%  
\section{Materials and Methods}

\subsection{Enrichment of Bacteria}

\begin{figure}
  \centering
\includegraphics[width=\linewidth]{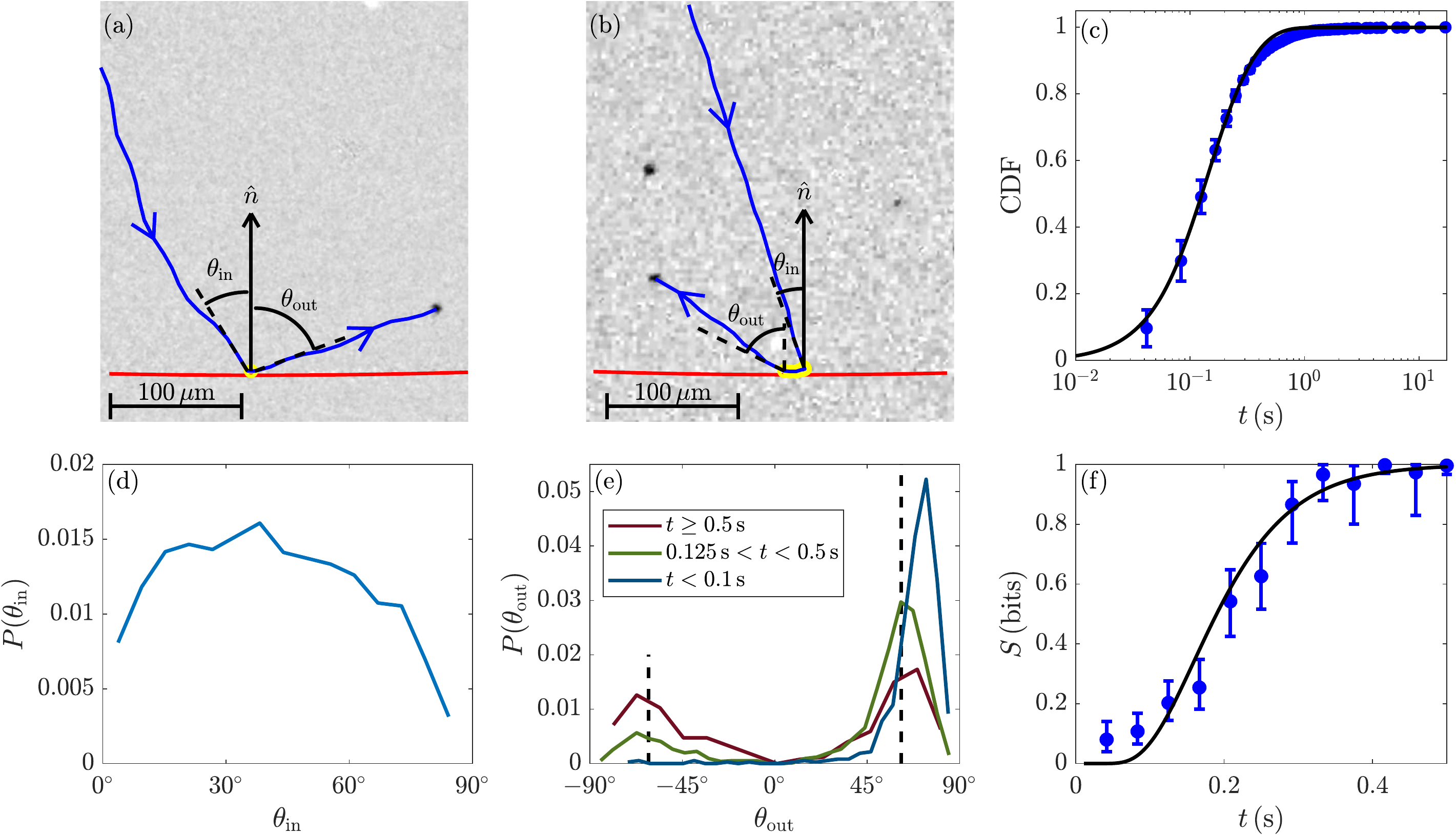}
\caption{When a \thio~cell collides with a hard wall it may become
  trapped by the surface. (a) A cell is only momentarily in contact
  with the surface (yellow dot). The blue line shows the trajectory of
  the cell with arrows indicating the direction of motion. The
  incident angle $\thin$ and escape angle $\theta_{\rm out}$ are
  measured from the surface normal $\hat{n}$. The coordinate system is
  chosen such that $\thin>0$ for all collisions. $\theta_{\rm out}>0$
  if the sign of the tangential velocity does not change. (b) A
  trapped cell may remain in contact with the surface for several
  seconds. In this example $\theta_{\rm out}<0$. (c) The cumulative
  probability distribution of escape times is compared to the
  distribution of first passage times calculated from the
  Fokker-Planke equation (Eq.~\ref{eq:fptime}). (d) The incident
  angles $\theta_{\rm in}$ are broadly distributed. (e) The
  probability density function of escape angles is sharply peaked for
  cells in fleeting contact with the surface. As cells remain in
  contact with the surface, a symmetric peak emerges corresponding to
  cells that reverse their motion tangent to the wall (as in panel
  [b]). The dashed lines show the prediction of Eq.(5). (d) The
  binomial entropy of the escape directions increases continuously as
  cells remain in contact with the boundary. The black line is
  equivalent to the fit in Fig.~4(c).
\label{entropy}}
\end{figure}

Because there are no known techniques to grow \Thio~in pure culture,
cells must be enriched from environmental samples using well
established
methods~\cite{de1961observations,marshall2012single,petroff2015fast,petroff2018nucleation}.
We collect sediment from a shallow tide pool in Little Sippewissett
Marsh ($41.5762^\circ\,$N, $70.6391^\circ\,$W), which is near Woods
Hole, Massachusettes. This sediment is stored in the lab in a
container covered in $15\,$cm of natural sea water. After three to
five days in the container, the sediment-water interface becomes
euxinic and a \thio~veil forms a few millimeters above the
sediment~\cite{jorgensen1983colorless,fenchel1998veil}, often between
strands of eel grass. Cells are collected from a fresh veil with a
$1\,$ml pipette and lightly mixed. After collection the veil reforms,
typically within a day.

\subsection{Microfluidic Device}
%  To observe the scattering and trapping of cells by a hard surface,
%  cells are loaded into a microfluidic chamber, which is built using
%  .
%
Microfluidic chambers---which are produced using standard soft
photolithography techniques~\cite{whitesides2001soft}---are composed
from Polydimethylsiloxane (PDMS) and sealed on one side by a glass
slide. Chambers are quasi-two dimensional, with a height of both
$150\,\mu$m and centimeter-scale lateral dimensions.
Two similarly designed microfluidic chambers are used in these
experiment.
In the first set of experiments the shape of the chamber, when viewed
from above, is a square ($1\,\mbox{cm}\times1\,\mbox{cm}$). 
In the second set of experiments (as in Fig.~1[a,b]), the chamber is
circular with a radius of $0.75\,$cm.
We initially suspected that the slight curvature of the wall could
lead to measurable differences in the scattering dynamics, however the
statistics presented here were found to be indistinguishable between
the two chamber designs. 
Because we limit our analysis to the trajectories of cells within
$80\,\mu$m of a wall, the curvature of a walls is small, being within
$1\%$ of flat.
Consequently, these measurements are combined in the analysis
presented here.

\subsection{Tracking of Cells}
\thio~cells are inoculated into the microfluidic chambers describe
above. The motion of cells as they swim near the outer wall of the
chamber is observed using the Zeiss $20\times$ objective, which is
focused to a plane $75\,\mu$m between to top and bottom of the
chamber to limit hydrodynamic interactions between the swimming cell
and the chamber.
In each experiment, the motion of cells near one wall of the chamber
is recorded at $24\,$fps using a Nikon D7000 camera.

We track the motion of cells swimming within $80\,\mu$m of the wall of
the chamber.
Swimming cells appear as dark spots under transmitted light. To
identify cells, we first average all frames over the course of a five
minute experiment. This background image is subtracted from each frame
of the video to highlight motion.
Swimming cells are identified using an intensity threshold.
These instantaneous measurements of cell position are connected into
trajectories by applying Munkres’ Assignment Algorithm.
Two representative trajectories are shown as solid blue lines in
Fig~1(a,b).
A small fraction of the trajectories corresponded to cells that are
close to division. These cells are atypically large and swim in
slightly helical trajectories.
To better distinguish torques arising from interactions with the
chamber walls from those that are due to the particular flagellation
pattern of the cell these helical trajectories are discarded.

From each trajectory, we calculate the velocity $\mathbf{v}$ of each
cell that collides with the wall of the chamber.
Cells are identified as in contact with the surface if the distance
between the center of the cell and the chamber wall is less than or
equal to the measured radius of the cell.
To find the incident angle $\theta_{\rm in}$, we measure the incident
velocity $\mathbf{v}_{\rm in}$ from the frames immediately before the
cell contacts the surface and measure angle between $\mathbf{v}_{\rm
  in}$ and the local normal $\hat{n}$ at the point of first contact.
Similarly, the asymptotic escape angle $\theta_{\rm out}$ is measured
relative to the local surface normal at the point where the cell
escapes. The velocity of the escaping cell $\mathbf{v}_{\rm out}$ is
calculated when the cell is two body lengths ($z=4a$) from the
surface, at which point hydrodynamic torques are negligible (see
Fig~2). The slight curvature of the circular chamber wall creates an
ambiguity in $\theta_{\rm out}$ of less than $0.2^\circ$, which we
ignore.

\begin{figure}
\centering \includegraphics[width=.5\linewidth]{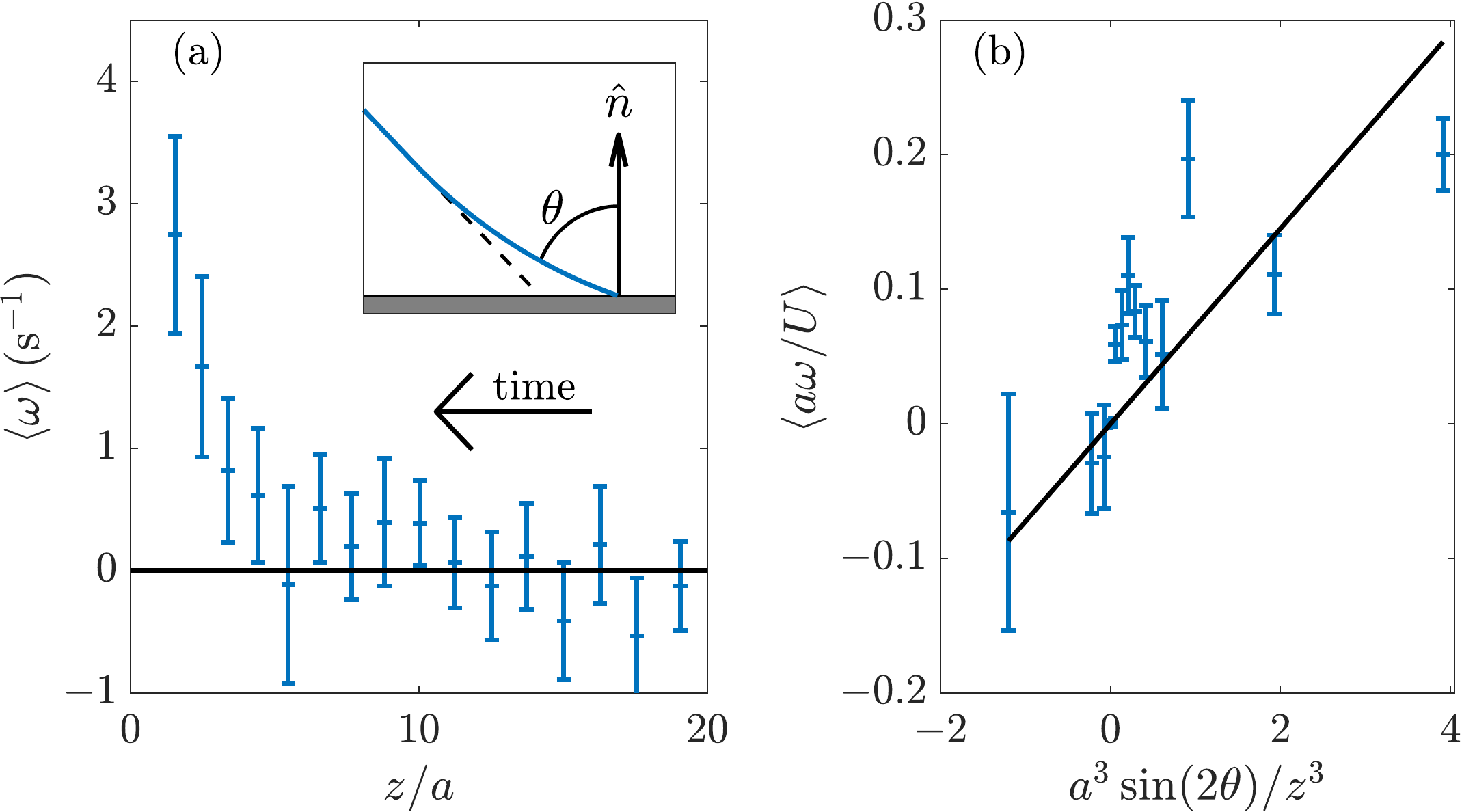}
\caption{Cells approaching a wall turn parallel to the surface. (a)
  When $z\gg a$, cells move swim in straight trajectories. When
  $z\lesssim 2a$ it begins to turn parallel to the surface. (inset)
  Schematic illustrating the trajectory (blue) of a cell as it strikes
  the wall. The dashed line shows a straight trajectory. (b) The
  measured rate that cells rotate is compared to the predictions of
  the far-field hydrodynamic predictions Eq.~(\ref{eq:far}). Only
  cells that are more than one body length from the wall are included in
  these panels.
\label{turning}}
\end{figure}

\section{Results}

\subsection{Motion of a cell far from a wall}
%% Approach to the wall
We begin by analyzing a cell's approach to a surface.
Fig.~\ref{turning}(a) shows that the average angular velocity of cells
$\langle\omega\rangle$ decreases with the distance $z$
between the cell and the wall.
Cells initially approach the wall at a constant angle,
$\langle\omega\rangle\approx0$.
However, when the cell arrives to within a distance of $z_c\sim 4 a$,
it begins to turn parallel to the
surface.%~$\langle\partial_t\theta_{\rm}\rangle>0$.

These dynamics are explained by the far-field hydrodynamics of a
pusher cell near a
wall~\cite{lauga2009hydrodynamics,drescher2011fluid}.
At distances $z\gg a$, the cell can be approximated as a force dipole.
A swimming cell moves as it pushes on the fluid and is advected and
rotated by its hydrodynamic image~\cite{blake1971note,lauga2009hydrodynamics}.
As \thio~is well approximated as a sphere~\cite{schulz2001big}, the
predicted angular velocity is~\cite{drescher2011fluid}
\begin{equation}
  \omega_{\rm far}=\dfrac{U}{a}\dfrac{9\ell a^3}{64 z^3}\sin(2\theta),
\label{eq:far}
\end{equation}
where $\ell$ is ratio of the dipole length of the cell to its radius.
% Note that the trajectory of a cell swimming directly towards the
% surface $\theta_{\rm in}=0$ is an unstable fixed point.
%  
Fig~\ref{turning}(b) shows fair agreement between the theory and
observation with a single fit parameter $\ell=0.47 \pm 0.18$.

\subsection{Motion of a cell near a wall}
The observation that cells align their flagella with the surface
normal when they are in contact with a boundary (see
Fig~\ref{entropy}[b], supplemental video S2, and
Refs.~\cite{petroff2018nucleation,das2019transition,ishimoto2019bacterial})
but not as they approach the boundary (see Fig.~\ref{turning})
strongly constrains the functional form of the torque acting on a
cell.
As the fixed point $\theta=0$ changes its stability, this bifurcation
must be either a transcritical bifurcation or a pitchfork
bifurcation~\cite{strogatz2018nonlinear}. The former of these options
is excluded by the symmetry of the system, which requires that the
angular velocity $\omega_{\rm near}$ be an odd function of $\theta$.

\begin{figure}
  \centering
\includegraphics[width=.333\linewidth]{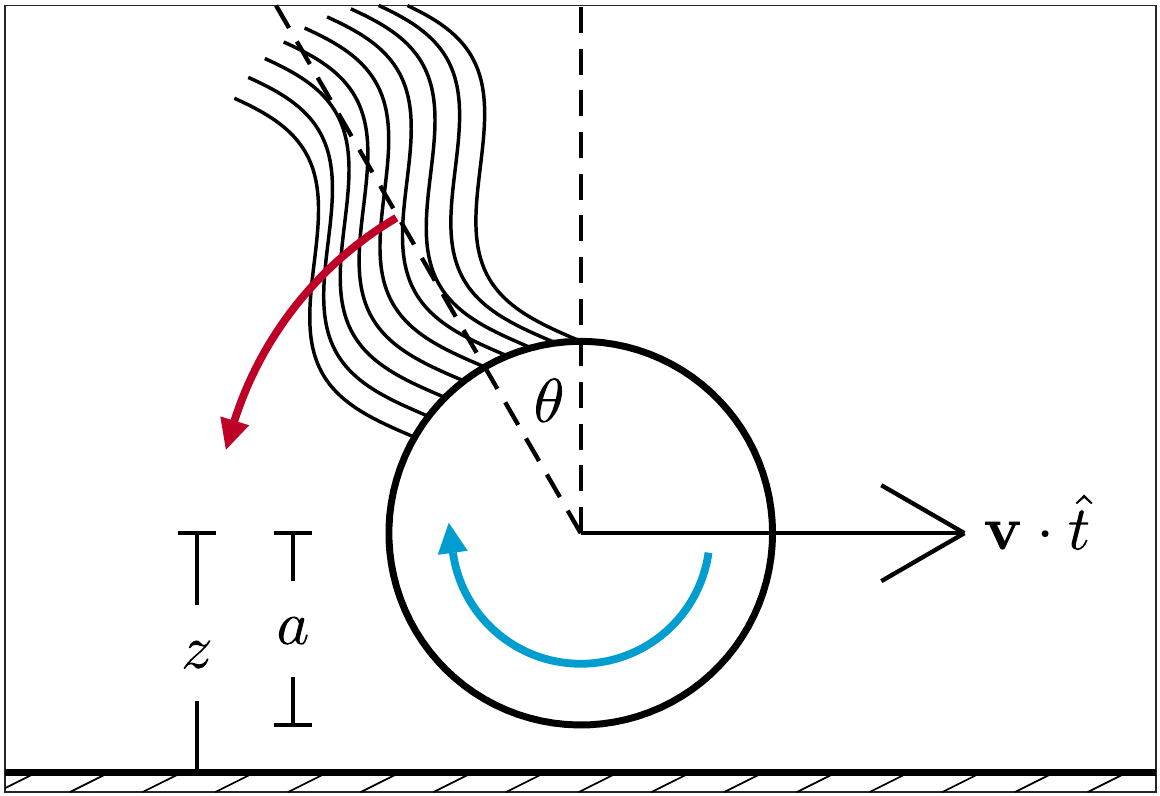}
\caption{As a cell moves laterally over a surface at speed
  $\mathbf{v}\cdot\hat{t}$, it is turned by two torques. Because
  velocity gradients in the gap between the cell and the surface are
  sharper than those between the cell the bulk fluid, there is a
  torque (blue arrow) that turns the flagella normal to the
  surface. This torque is countered the torque on the flagella. Both
  drag on the flagella and the hydrodynamic attraction of the flagella
  to the wall turn (red arrow) the flagella parallel to the
  surface. Because the flagella of \thio~are relatively short,
  $\theta=0$ is a stable fixed point~\cite{das2019transition}. This
  schematic is drawn close to the proposed unstable orientation at
  which the two torques balance.
\label{cartoon}}
\end{figure}

As illustrated in Fig~\ref{cartoon}, physical reasoning suggests that
the orientation $\theta=0$ undergoes a subcritical pitchfork
bifurcation as the cell approaches the wall.
Consider the torques acting on a cell in contact with a wall.
If the flagella are slightly misaligned with the local normal, the
cell moves laterally over the surface.
As the resulting velocity gradients are greater in the narrow gap
between cell and the surface than between the top of the cell and the
bulk fluid, there is a net torque on the cell that turns the flagella
normal to the surface~\cite{petroff2018nucleation,das2019transition}.
This torque is countered by drag on the flagella, however because the
flagella of \thio~are relatively short compared to the cell size,
viscous dissipation between the cell and the wall dominates, and
$\theta=0$ is stable~\cite{das2019transition}.
As the angle $\theta$ between the flagella and the surface normal
grows, the flagella are rotated closer to the surface and hydrodynamic
attraction~\cite{das2019transition} between the flagella and the surface becomes dominant.
These competing torques suggest there is an unstable orientation
$\theta_{\rm b}$ at which the torque due to the flow between the cell
body and the surface balance hydrodynamic attraction between the
flagella and the surface.
Thus, we expect that as the cell approaches the wall, the fixed point
$\theta=0$ undergoes a bifurcation that stabilizes this orientation
while generating unstable fixed points at $\pm\theta_{\rm b}$. This
process is typical of a subcritical pitchfork bifurcation.

This reasoning gives little insight into torques acting on cell that
is turned to large angle, where contact forces between hundreds of
rotating flagella and the surface are likely important.
In response to this uncertainty, we make the simple assumption that
the angular velocity saturates to some maximum value as $\omega_{\rm
  near}(\theta\gg\theta_{\rm b})=\omega_{\rm max}$.

To analyze the motion of cells in contact with a surface, we consider
a minimal model that includes the normal form a subcritical pitchfork
bifurcation predicts a bounded angular velocity.
We propose 
\begin{equation}
  \omega_{\rm near}(\theta)=
\begin{cases}

  K\left( \left(\dfrac{\theta}{\theta_{\rm b}}\right)^2-1\right)\theta &
  |\theta|\leq\theta_{\rm c}\\

    \mbox{sign}(\theta)\omega_{\rm max} & |\theta|>\theta_{\rm c}\\

\end{cases}
\label{eq:near}
\end{equation}
where the angle $\theta_{\rm b}$ defines the edge of the basin of
attraction of the bound orientation $\theta=0$, $K$ is a rate
coefficient, and $\theta_{\rm c}$ is chosen such that $\omega_{\rm
  near}$ is continues.

% A cell collides with a surface at an angle $\theta_{\rm in}$ and is
% rotated by the hydrodynamic and Brownian torques, which are
% characterized by a rotational diffusion coeficient $D_{\rot}$. If
% these torques rotate the cell's orientation $\theta$ to within the
% basin of attraction before it reaches $\theta=0$, the cell esacpes at
% an angle $\theta_{\rm out}>0$. If the cell first reaches $\theta=0$,
% its eventual escape is symmetric and the entropy $S$ is maximized.
% %
% This reasoning maps the scattering and trapping of cells to a first
% passage problem.% by analyzing a Focker-Planck Equation.

To test this hypothesis, we map the escape of a cell from the wall
onto a first passage problem.
Our measurements provide the probability $P_+$ that a cell that strikes
the surface at a particular angle $\thin$ escapes at an angle
$\theta_{\rm out}>0$ within a time $t$.
We recapitulate these measurements in our model.
For a given choice of $\theta_{\rm b}$, $K$, $\omega_{\rm max}$, and
rotational diffusion coefficient $D_{\rm rot}$, we calculate the
probability that a cell is rotated to an orientation $\theta=90^\circ$
(at which it escapes in the positive sense) before it is turned to an
orientation $-90^\circ$.
We calculate how this probability varies with the incident angle and
find the distribution of first passage times.

%  This model is sufficient to qualitatively understand the observed
%  differences in cell motion (see Fig.~\ref{entropy}).
%  %
%  After a collision, a cell is reoriented by hydrodynamic torques
%  according to Eq.~(\ref{eq:near}) and Brownian torques, which are
%  characterized by a rotational diffusion coeficient $D_{\rm rot}$.
%  %
%  When these torques orient the cell to an angle of $\pm90^\circ$, the
%  cell escapes into the bulk fluid. % the positive sense (Fig.~\ref{entropy}[a]).
%  %
%  Cells that are captured by the stable fixed point are equally likely
%  to eventually escape in either direction.
%  %
%  For a given choice of the model parameters and initial conditions, we
%  calcuate the probability $P_+$ that a cell escapes in the positive
%  sense.

We solve this first passage problem by way of a Fokker-Planck
equation.
Let $p(\theta,t)$ be the probability density that the flagella of cell
exert a net force oriented at an angle $\theta$ off of the surface
normal at time $t$.
The probability distribution evolves in time as
\begin{equation}
  \dfrac{\partial p}{\partial t}=-\dfrac{\partial}{\partial
    \theta}\left(\omega_{\rm near} p\right)+D_{\rm
    rot}\dfrac{\partial^2 p}{\partial \theta^2}.
  \label{eq:fptime}
\end{equation}
For a given choice of model parameters, we discretize the interval
$-\pi/2<\theta<\pi/2$ into 300 elements and numerically integrate this
equation using standard finite element methods for linear
equations~\cite{reddy2005introduction}.
We then calculate the probability flux $j=\omega_{\rm near}p-D_{\rm
  rot}\partial p/\partial\theta$. The probability that a cell escapes
at a positive angle (as in Fig.~\ref{entropy}[a]) is
$P_+=j_+/(j_++j_-)$, where $j_\pm$ is the probability flux evaluated
at the absorbing boundaries $\theta=\pm\pi/2$.
%  , which correspond to the escape of cells into the
%  bulk fluid. 

\begin{figure}[t]
  \centering
\includegraphics[width=.5\linewidth]{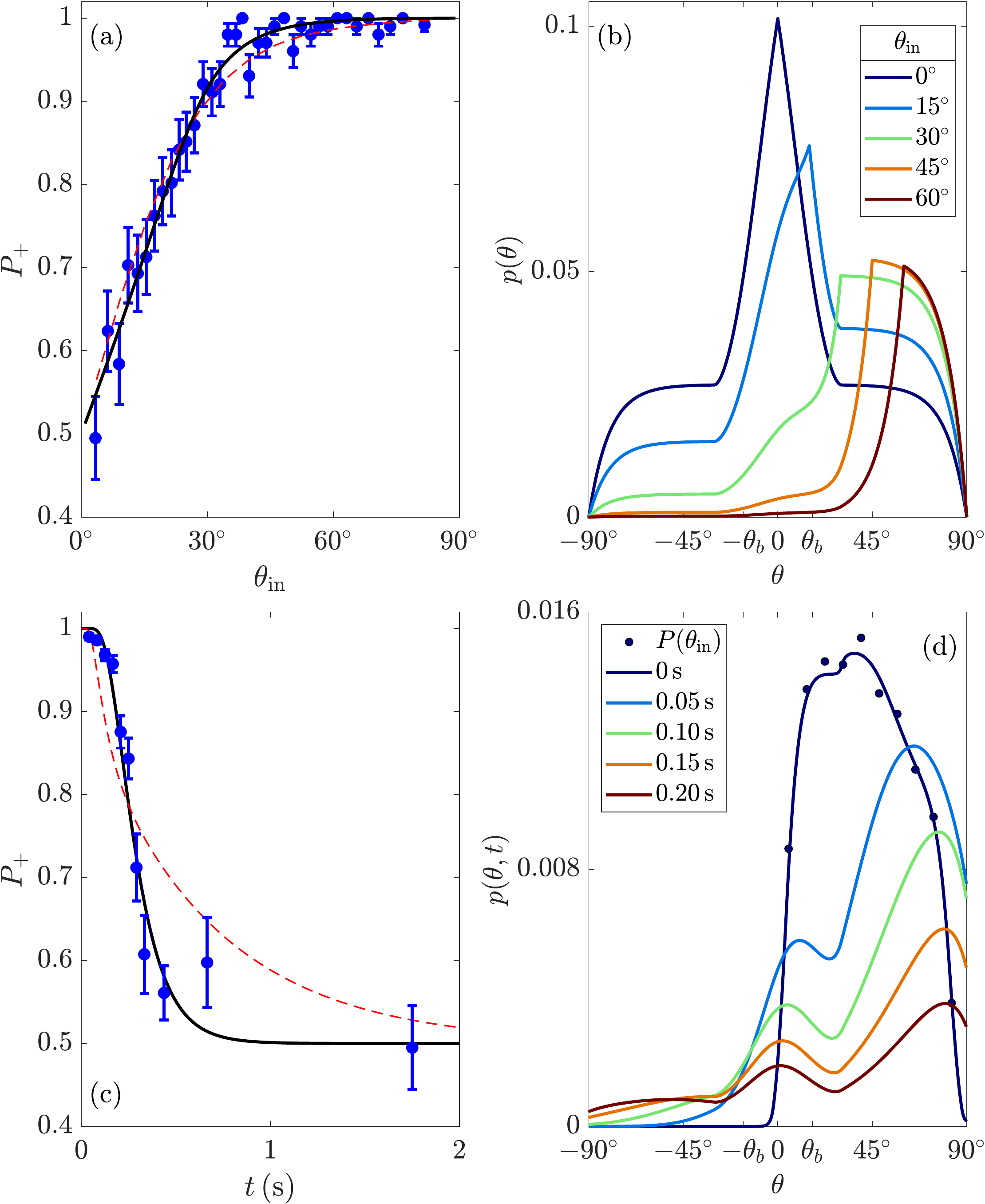}
\caption{The probability $P_+$ that a cell escapes the surface at an
  angle $\theta_{\rm out}>0$ is fit to Eq.~(3). (a) $P_+$ is measured
  for several thousand collisions at a variety of incident angles
  $\thin$. The black line shows the best fit. (b) The steady state
  distribution of cells are shown with sources at several choices of
  $\thin$. These distributions are normalized such that
  $j_++j_-=1$. (c) Cells that escape the surface quickly escape at
  positive angles ($P_+=1$). Those that become trapped eventually
  escape with equal probability in either direction ($P_+=1/2$). The
  black line shows the best fit to the Fokker-Planck equation. These
  data are the same as in Fig.1(d) but scaled linearly to better
  compare to the model. (d) Solutions of the Fokker-Planck equation
  are shown at five time points. The initial distribution is
  interpolated from the measured distribution of incident angles
  $P(\theta_{\rm in})$.
  \label{trapping_th}}
\end{figure}

We first consider the fate of a cell that collides with the surface at
an angle $\theta_{\rm in}$ (see Fig~\ref{trapping_th}[a]).
To find the fraction of cells that eventually escape in the positive
sense, we solve for the Greens functions of the steady-state
Fokker-Planck equation with a source of cells at $\thin$.
%  \begin{equation}
%    0=-\dfrac{\partial}{\partial
%      \theta}\left(\dfrac{\omega_{\rm near}}{D_{\rm rot}}
%    p\right)+\dfrac{\partial^2 p}{\partial
%      \theta^2}+\delta(\theta-\thin),
%    \label{eq:fpgreen}
%  \end{equation}
% where $\delta(\theta)$ is the Dirac delta function.
%
%  We numerically integrate to find the fluxes at the absorbing
%  boundaries $\theta=\pm 90^\circ$ for sources of cells at each of the
%  discritized values of $\theta$.
%
Figure~\ref{trapping_th}(b) shows solution for five values of $\thin$.
Note that as $\thin$ increases, the fraction of cells that accumulate
at $\theta=0$ decreases.

We fit three parameters to match these solutions to the measured
variation of $P_+$ (see Fig~\ref{trapping_th}[a]).
The first parameter is the width of the basin of attraction of trapped
orientation $\theta=0$. We find that $\theta_{\rm b}=15^\circ \pm
2^\circ$.
%
%  This value is consistent with a previous independent estimate of
%  $\theta_{\rm b}\sim10^{\circ}$.
%
%The other two fit parameters correspond to the rotational Peclet
%numbers of cells~\cite{note}.
%
The second fit parameter $\sqrt{D_{\rm rot}/2K}=22^\circ\pm5^\circ$
represents the typical fluctuations about the stable orientation
$\theta=0$.
Because these fluctuations are somewhat larger than the basin of
attraction of this fixed point, we conclude that cells in this
experiment are only momentarily bound.
%  
%  first of these Peclet numbers $\mbox{Pe}_0=\theta_{\rm b}^2
%  K/D_{\rm rot}=0.08\pm0.04$ is the typical ratio of hydrodynamic
%  torques to Brownian torques for a hydrodynamically bound cell.
%  %
%  That this value is reasonably small indicates that cells quickly
%  escape from this fixed point.
%
The final fit parameter $\mbox{Pe}=\pi \omega_{\rm max}/2D_{\rm
  rot}=5.7\pm2.5$ is the rotational Peclet number, which represents
the ratio of the rates at which a cell is turned by hydrodynamic and
Brownian torques.
Its reasonably large value indicates that cells that collide with the
surface outside the basin of attraction of $\theta=0$ rarely become
trapped.

Next, we consider how the probability that a cell escapes in the
positive sense decays as it remains in contact with the surface, as
shown in Fig.~\ref{trapping_th}(c).
We solve Eq.~(\ref{eq:fptime}) using the measured distribution of
$\thin$ as the initial distribution of orientations and the parameter
values fit to Fig.~\ref{trapping_th}(a).
Figure~\ref{trapping_th}[d] shows the evolution of distribution of
orientations as cells escape from the surface and become trapped at
$\theta=0$.
These dynamics uniquely define the functional form of $P_+(t/\tau_{\rm
  D})$, where the diffusive timescale $\tau_{\rm D}=\pi^2/D_{\rm rot}$
is the only unknown parameter.
We fit $\tau_{\rm D}$ by rescaling the measured decay of $P_+$ to the
predicted functional form.
We find $D_{\rm rot}=1.15\pm0.1\,$rad$^2$/s, which implies $K=4.22 \pm
2.11\,$s$^{-1}$ and $\omega_{\rm max}=6.92 \pm 0.18\,$rad/s.
Fig.~1(c) compares the measured cumulative distribution function of
escape times to the probability $\int_0^Tj_+(t) +j_-(t)\mathrm{d}t$
that a cell escapes before time $T$.
Similarly, the escape entropy corresponding to the best fit of
$P_+(t)$ is shown in Fig.~1(f).

%  The magnitudes of these values are qualitatively consistent with the
%  directly measured angular velocity of cells as they approach the
%  surface.

The measured variations of $P_+$ with incident angle and time are
inconsistent with the physical null model that the far field
hydrodynamic effect of a boundary can be extended to model the motion
of a cell in contact with a surface.
We repeat the procedure described above to fit $P_+$ to the
Fokker-Planck equation where the drift velocity is described by
Eq.~\ref{eq:far} and the prefactor on the sinusoidal term is fit.
As shown in Fig.~3(a--b), while the variation with incident angle is
equally well described by either model, the first passage times differ
markedly.

\begin{figure}
\centering
\includegraphics[width=.5\linewidth]{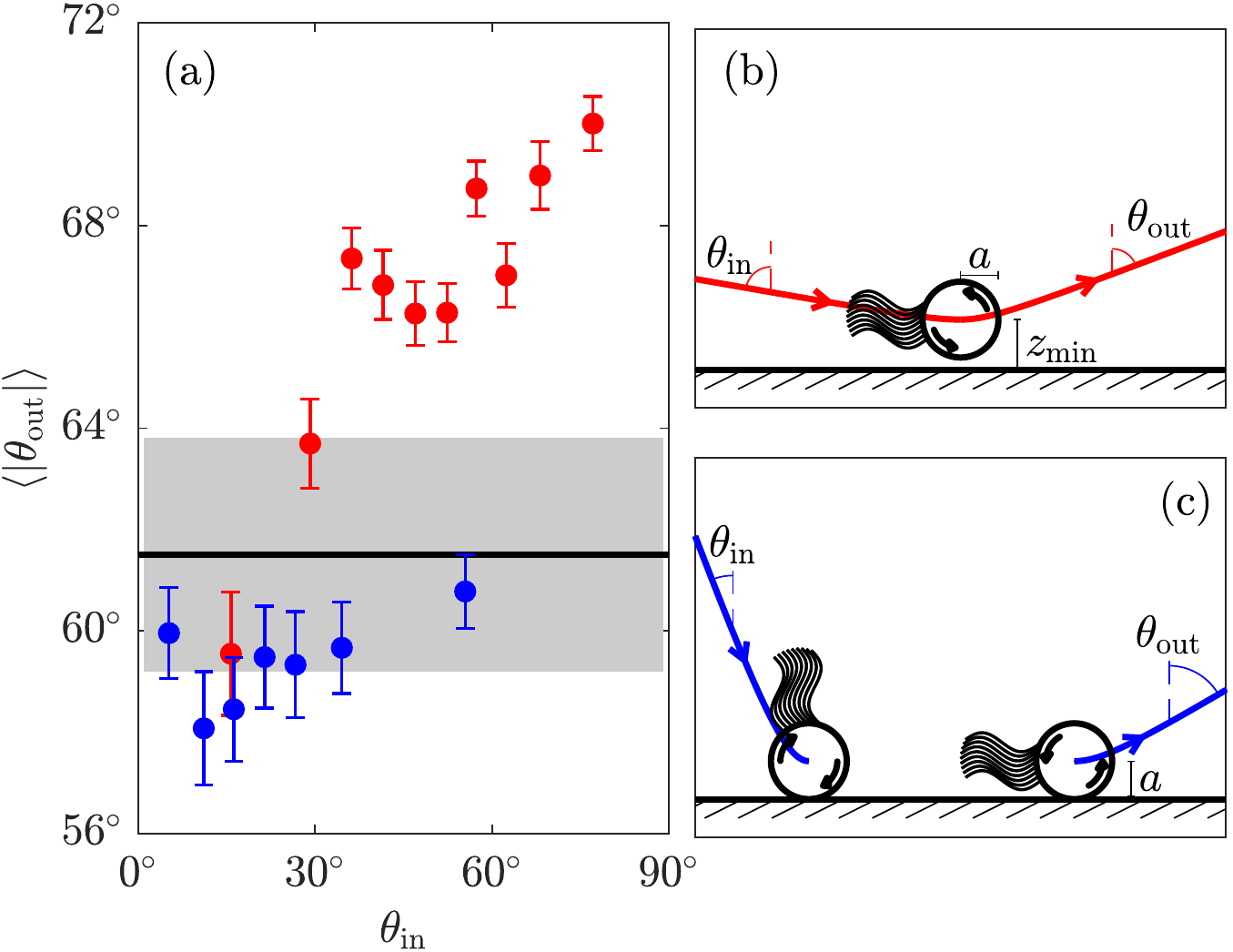}
\caption{The escape angle is determined by the minimum distance
  between the cell and the wall. (a) The average escape angles (blue
  dots) of cells that remain in contact for at least $0.125\,$s are
  similar with the prediction (dashed line) of Eq.~(5). Cells that
  remain in contact with the wall for shorter durations escape at
  slightly greater angles (red dots). The colors of these dots
  correspond to the conjectured trajectories shown in the first two
  panels.  (b) The red line shows the trajectory of a cell that
  approach the surface at shallow angles and is turned by hydrodynamic
  interactions with the surface before they collide. They escape from
  a height $z_{\rm min}$. (c) Cells that collide with the surface at
  sharper angles eventually escape from a height of a cell radius
  $a$. The blue line illustrates the approach and eventual escape of a
  of a cell that is briefly trapped.
  \label{fig:4}}
\end{figure}

\subsection{Matching near and far field dynamics}
The agreement between theory and observation shown in Fig.~3 leads us
to conclude that Eq.~(2) captures the essential qualities of the near
field dynamic coupling between a \thio~cell and a wall.
Two questions remain. First, these results do not constrain the
distance between a cell and the wall at which a cell may become
trapped.
Additionally, it is not clear why all cells escape the surface at
narrowly distributed angles (see Fig.~1[e]).

To answer these questions we match the short-range angular velocity
(Eq.~[2]) with the far field (Eq.~[1]).
To match the cubic decay of torques on the cell predicted by the far
field dynamics, we propose that at short distances $z\sim a$ from the wall
and small angles ($\theta<\theta_{\rm c}$)
\begin{equation}
  \omega(z,\theta)= \dfrac{K \theta}{((z-a)/\eta +1)^4}\left(
  \left(\dfrac{\theta}{\theta_{\rm b}}\right)^2+\dfrac{z-a}{\eta}-1\right),
\end{equation}
where $\eta$ is the distance from the wall at which the orientation
$\theta=0$ becomes stable.
In the limit $z\gg\eta$ and $\theta\ll\theta_{\rm b}$, this expansion
matches Eq.~(1) where $\eta/a=(9\ell U/32 Ka)^{1/3}=1.27\pm0.16$.
We conclude that cells may become bound when the gap between the cell
and the wall becomes similar to the cell radius. This result is
intuitively consistent with the physical model summarized in
Fig.~\ref{cartoon}, in which the bound state is stabilized because
velocity gradients in the gap between cell and surface are sharper
than those between the cell and the bulk fluid.

%  This value of $\eta/a$ is our previously described
%  model for the stability of this bound state. We conjectured that, as
%  the cell moves laterally over the surface, velocity gradients in the
%  gap between the cell and the wall exceed those between the cell and
%  bulk fluid. The gradient of viscous forces across the cell causes the
%  cell to roll and align its flagella with the surface normal. This
%  effect should only become dynamically important when the distance
%  between the cell and surface becomes similar to the cell size (i.e.,
%  $\eta/a\approx1$).

Finally, we consider the escape of a cell from the surface.
All cells escape by first swimming tangent to the wall.
We consider the simplest function that interpolates between the
constant angular velocity $\omega_{\rm max}$ found when the cell is in
contact with the surface the cubic decay predicted by the far
field. We expect the angular velocity of an escaping cell to decay as
\begin{equation}
  \omega(\theta,z)=\dfrac{\omega_{\rm max}}{((z-a)/\eta+1)^3}.
\end{equation}
Cells escape with speed $\dee z/\dee t=-U\cos(\theta)+u_{\rm
  im}(z,\theta)$, where $u_{\rm im}(z,\theta)$ is a correction to the
swimming speed due to advection by the cell's hydrodynamic image. Its
functional form is provided in Ref.~\cite{drescher2011fluid}.

We numerically integrate Eq.~(5) to find the asymptotic orientation of
a cell's motion that is initially swimming tangent to the surface.
Taking the initial distance between the cell and the wall to be one
cell radius, we find---with no fit parameters---a predicted escape
angle of $61.5^\circ \pm 2.3^\circ$.
As shown in Fig.~4, this value is consistent with the measured escape
angle $\langle|\theta_{\rm out}|\rangle=59.5^\circ \pm 0.7^\circ$
found for cells that remain in contact with the surface for at least
$0.125\,$s.
Cells that are identified by the tracking algorithm to be in contact
with the surface for shorter durations escape at slightly greater
angles, which increase with the incident angle.
We believe that these cells are turned away from the surface by
hydrodynamic torques before they come in physical contact with the
wall, similar to the scattering of
\textit{Chlamydomonas}~\cite{lushi2017scattering}.
Taking the minimum distance between the cell and wall to be $z_{\rm
  min}$, the observed increase in $\langle |\theta_{\rm out}| \rangle$
is consistent with a value of $z_{\rm min}$ that increases
monotonically with $\thin$.

%  Cells that are identified by the tracking algorithm to be in contact
%  with the surface for a single frame escape at a slightly greater angle
%  of $\langle|\theta_{\rm out}|\rangle=69.2^\circ \pm 0.8^\circ$.
%  %
%  We suspect that these cells are turned away from the surface by
%  hydrodynamic interactions before they make physical contact.
%  %
%  The average escape angle of these cells is consistant with
%  trajectories that approach to a minimum distance of $z_{\rm
%    min}/a=1.33\pm0.04$.
%  %
%  The increase of escape angles with incident angle for cells that are
%  trapped for intermediate times is consistant with a value of $z_{\rm
%    min}$ that increases linearly $\thin$.

%short: theta_out=
%long: theta_out=60.8 pm 0.6

%  To understand the monotonic increase of $\langle|\theta_{\rm
%    out}|\rangle$ with $\thin$, note that a cell that approches a
%  surface along the tangent may by turned away from the surface by
%  hydrodynamic interactions before making physical contact.
%  %

%  With no additional fit parameters, we predict that cells escape the
%  surface at an angle $70^\circ \pm 1^\circ$ (see
%  Fig.~\ref{entropy}[d]). This value agrees strikingly well with the
%  measured escape angle $|\theta_{\rm out}|=69^\circ \pm 3^\circ$.

\section{Conclusion}
In conclusion, we have used the scattering statistics of collisions
between \thio~and a hard wall to probe the near-field dynamics between
a fast-swimming multiflagellated cell and a surface.
A simple physical picture emerges. As a cell approaches a wall,
dipole-dipole interactions with its hydrodynamic image turn the cell
to swim parallel to the surface.
When the distance between the cell and the surface decreases to a value
of $1.27$ cell radii the torques acting on the cell change
qualitatively.
Shear forces in the gap between the cell and the surface orient the
cell to exert a force normal to the surface. The cell becomes trapped.
We find that the basin of attraction of the fixed point is narrow.
Consequently, most cells collide with the surface outside of this
basin and rapidly escape while maintaining their direction of motion
tangent to the surface.
When a cell is captured by the stable fixed point, its eventual
escape is symmetric and all information of its approach to the wall is
erased.
Because all cells that come in contact with the surface escape by
first swimming tangent to the wall at a distance of one cell radius,
all cells escape at similar angles.
A minority of cell are turned from the wall before coming in physical
contact with the wall and escape at slightly greater angles.

Because the near field dynamics described in Eq.~(2) are
phenomenological, it remains unclear whether the stability of trapped
cells is primarily due to hydrodynamic of contact forces.
We suspect hydrodynamic torques dominate at small $\theta$ and
collisions become increasingly important as the flagella are turned
toward the wall.
Indeed, Das and Lauga~\cite{das2019transition} analyzed a simplified
model of \thio~with a single short flagellum to show that hydrodynamic
torques are sufficient to stabilize the bound state. However, this
analytic model---which was used primarily to provide intuition for an
aspect of their more detailed numeric results---neglected hydrodynamic
attraction of the flagella to the surface.
As the simplified cell approaches the surface, the orientation
$\theta=0$ becomes stable through a supercritical pitchfork
bifurcation, rather than the subcritical bifurcation found here.
We encourage computational fluid mechanicians to simulate the fluid
flow around a spherical cell with a single flagellum near a wall and
compare the angular velocity of a cell to Eqs.~(2) and its matching to
the far field.

The trapping of \thio~by a hard surface is the first step in the
nucleation of active chiral crystals\cite{petroff2015fast}.
In our previous study, we found that isolated cells remain trapped by
the surface for tens of seconds~\cite{petroff2018nucleation}, much
longer than average trapping time $0.21\,$s found here.
We ascribe this difference to the roughness of the walls made of PDMS
to those made of polished glass.
As the gap between the cell and the surface is presumably much smaller
if the surface is smooth, the velocity gradient between the cell and
the wall---which stabilize the trapped cell---is much sharper.
Consequently, we expect cells by be more strongly trapped by cover
slips than microfluidic walls.
The escape of cells could be further enhanced by the sporadic binding
of quickly-rotating flagella to the PDMS, which may increase the
effective rotational diffusion of cells, causing cells to escape more
quickly from PDMS than from glass.
This result highlights the importance of an unnaturally smooth surface
for the formation of active chiral crystals. We are consequently
doubtful of the biological significance of this form of collective
motion.

Nonetheless, the dynamics by which cells scatter from rough surfaces
are likely quite important for the ecology of \thio.
These bacteria, which live in the pore space of water saturated sand,
exude mucus tethers from their posteriors to attach to sand
grains~\cite{schulz2001big}.
It is not understood how cells attach this mucus thread to a surface.
Our results show that when a cell collides with a surface outside of a
narrow basin of attraction, it rapidly escapes rather than swimming
parallel to the surface.
As free-swimming \thio~frequently drag short tethers as they
swim~\cite{fenchel1998veil} and collisions quickly turn the cell
posterior toward the wall, it is plausible that these collisions
facilitate attachment of the tether to the surface.
Given a swimming speed of $600\,\mu$m/s and a pore size of several tens
of micron, these collision dynamics provide several opportunities a
second for a dragged tether to stick to a surface.

\begin{acknowledgments}

We thank Dabasish Das and Arshad Kudrolli for their insightful
comments and discussions. This work was supported by NSF Grant
no. PHY-2042150.
\end{acknowledgments}

\appendix
\section{Appendix: Supplemental Videos}
Two videos of collisions between a cell and a surface are provided.
\begin{enumerate}
\item Video SV1.mp4 shows a representative collision of a \thio~with
  the wall of a microfluidic chamber in which the cell is only
  momentarily in contact with the surface. The trajectory is similar
  to that shown in Fig.~1(a).
\item SV2.mp4 shows a collision between the cell that strikes the
  surface close to the local normal and becomes trapped. This
  collision is also shown in Fig~1(b).
\end{enumerate}

\bibliographystyle{unsrt}
% \bibliography{thio}

\end{document}